\documentstyle[psfig]{mn}

\newif\ifAMStwofonts
\newcommand{\lapprox}{\stackrel{<}{\scriptstyle \sim}}
\newcommand{\gapprox}{\stackrel{>}{\scriptstyle \sim}}

\ifoldfss
  \ifCUPmtlplainloaded \else
    \NewTextAlphabet{textbfit} {cmbxti10} {}
    \NewTextAlphabet{textbfss} {cmssbx10} {}
    \NewMathAlphabet{mathbfit} {cmbxti10} {} % for math mode
    \NewMathAlphabet{mathbfss} {cmssbx10} {} %  "   "    "
  \fi
  \ifAMStwofonts
    \ifCUPmtlplainloaded \else
      \NewSymbolFont{upmath} {eurm10}
      \NewSymbolFont{AMSa} {msam10}
      \NewMathSymbol{\upi}     {0}{upmath}{19}
      \NewMathSymbol{\umu}     {0}{upmath}{16}
      \NewMathSymbol{\upartial}{0}{upmath}{40}
      \NewMathSymbol{\leqslant}{3}{AMSa}{36}
      \NewMathSymbol{\geqslant}{3}{AMSa}{3E}

       \let\le=\leqslant
       \let\ge=\geqslant
    \fi
  \fi
\fi % End of OFSS

\ifnfssone
  \newmathalphabet{\mathit}
  \addtoversion{normal}{\mathit}{cmr}{m}{it}
  \addtoversion{bold}{\mathit}{cmr}{bx}{it}
  \newmathalphabet{\mathbfit} % math mode version of \textbfit{..}
  \addtoversion{normal}{\mathbfit}{cmr}{bx}{it}
  \addtoversion{bold}{\mathbfit}{cmr}{bx}{it}
  \newmathalphabet{\mathbfss} % math mode version of \textbfss{..}
  \addtoversion{normal}{\mathbfss}{cmss}{bx}{n}
  \addtoversion{bold}{\mathbfss}{cmss}{bx}{n}
  \ifAMStwofonts
    \ifCUPmtlplainloaded \else
      \UseAMStwoboldmath
      \makeatletter
      \new@mathgroup\upmath@group
      \define@mathgroup\mv@normal\upmath@group{eur}{m}{n}
      \define@mathgroup\mv@bold\upmath@group{eur}{b}{n}
      \edef\UPM{\hexnumber\upmath@group}
      \new@mathgroup\amsa@group
      \define@mathgroup\mv@normal\amsa@group{msa}{m}{n}
      \define@mathgroup\mv@bold\amsa@group{msa}{m}{n}
      \edef\AMSa{\hexnumber\amsa@group}
      \makeatother
      \mathchardef\upi="0\UPM19
      \mathchardef\umu="0\UPM16
      \mathchardef\upartial="0\UPM40
      \mathchardef\leqslant="3\AMSa36
      \mathchardef\geqslant="3\AMSa3E

       \let\le=\leqslant
       \let\ge=\geqslant
    \fi
  \fi
\fi % End of NFSS release 1

\ifnfsstwo
  \DeclareMathAlphabet{\mathbfit}{OT1}{cmr}{bx}{it}
  \SetMathAlphabet\mathbfit{bold}{OT1}{cmr}{bx}{it}
  \DeclareMathAlphabet{\mathbfss}{OT1}{cmss}{bx}{n}
  \SetMathAlphabet\mathbfss{bold}{OT1}{cmss}{bx}{n}
  \ifAMStwofonts
    \ifCUPmtlplainloaded \else
      \DeclareSymbolFont{UPM}{U}{eur}{m}{n}
      \SetSymbolFont{UPM}{bold}{U}{eur}{b}{n}
      \DeclareSymbolFont{AMSa}{U}{msa}{m}{n}
      \DeclareMathSymbol{\upi}{0}{UPM}{"19}
      \DeclareMathSymbol{\umu}{0}{UPM}{"16}
      \DeclareMathSymbol{\upartial}{0}{UPM}{"40}
      \DeclareMathSymbol{\leqslant}{3}{AMSa}{"36}
      \DeclareMathSymbol{\geqslant}{3}{AMSa}{"3E}

       \let\le=\leqslant
       \let\ge=\geqslant
    \fi
  \fi
\fi % End of NFSS release 2

\ifCUPmtlplainloaded \else
  \ifAMStwofonts \else % If no AMS fonts
    \def\upi{\pi}
    \def\umu{\mu}
    \def\upartial{\partial}
  \fi
\fi

\title[Field Statistics for Vela's Pulsar: 3]
{Intrinsic Variability and Field Statistics for the Vela Pulsar: 3. 
Two-Component Fits and Detailed Assessment of Stochastic Growth Theory}
\author[Cairns, Das, Robinson, and Johnston] 
       {Iver~H. Cairns, P. Das, P.~A. Robinson, and S. Johnston \\
        School of Physics, University of Sydney, NSW 2006, Australia.}
\date{Accepted ??.
      Received 2002 ??;
      in original form 2002 ??}

\pagerange{\pageref{firstpage}--\pageref{lastpage}}
\pubyear{2002}

\begin{document}

\maketitle

\label{firstpage}

\begin{abstract}
The variability of the Vela pulsar (PSR B0833-45) corresponds to well-defined field 
statistics that vary with pulsar phase, ranging from Gaussian intensity statistics   
off-pulse to approximately power-law statistics in a transition region and then 
lognormal statistics on-pulse, excluding giant micropulses.     
These data are analyzed here in terms of two superposed wave populations, 
using a new calculation for the amplitude statistics of two vectorially-combined 
transverse fields. Detailed analyses show that the 
approximately  power-law and lognormal distributions observed are fitted well at 
essentially all on-pulse phases by  Gaussian-lognormal and double-lognormal 
combinations, respectively. The good fits found, plus the 
smooth but significant variations in fit parameters across the source, provide 
strong evidence that the approximately power-law statistics observed in the 
transition region are not intrinsic. Instead, the data are consistent with the 
pulsar's normal emission having lognormal statistics whenever the 
pulsar is detectable. This is consistent with 
generation in an inhomogeneous source obeying stochastic growth theory (SGT) and 
with the emission 
mechanism being purely linear (either direct or indirect), with 
no evidence for nonlinear processes. A nonlinear mechanism is viable only if 
it produces lognormal statistics when suitably ensemble-averaged. 
Variations in the SGT fit parameters with phase 
are consistent with the radiation being relatively more variable near the 
pulse edges than near the center, consistent with earlier work. In 
contrast, Vela's giant micropulses come from a very restricted phase 
range and have power-law statistics with indices ($6.7 \pm 0.6$) not inconsistent with 
nonlinear wave collapse. These results 
are consistent with normal pulses coming from a different source 
and generation mechanism than giant micropulses, as suggested 
previously on other grounds. Analysis of field statistics thus 
emphasizes the richness of pulsar physics, the apparently widespread 
applicability of SGT, and connections between variability and 
generation mechanism. 
\end{abstract}

\begin{keywords} 
pulsars: Vela; pulsars: general; radiation mechanisms: non-thermal; 
methods: statistical; waves; instabilities.
\end{keywords}

\section{Introduction}

   The variability of pulsar emissions, both from pulse to pulse at a given 
phase and from phase to phase within a pulse, has long been unexplained 
\cite{hankins1996}. These variations include subpulses \cite{drakecraft1968}, 
with durations of order $10 - 50\%$ the width of the average profile and 
sometimes a steady drift in phase, and microstructures superposed on the 
subpulses (Craft, Commella \& Drake 1968, Hankins 1996), which are concentrated bursts 
of emissions that sometimes appear quasi-periodic. 
The statistics of the variable fields, such as the probability 
distribution of fields, have not been characterized until recently (Cairns, Johnston 
\& Das 2001, hereafter Paper I) 
and few 
constraints have been placed on pulsar emission mechanisms, despite decades of 
research. This paper is the third in a series that characterizes the field statistics 
of pulsar variability in detail, advances interpretations in terms of existing theories for 
wave growth in inhomogeneous plasmas, and places constraints on pulsar emission 
mechanisms and the source physics. The initial analysis of Paper I, 
the second in the series \cite{paperII} [called paper II hereafter], and this paper 
address the Vela 
pulsar (PSR B8033-45), including both microstructure and subpulse effects 
simultaneously, while the fourth paper addresses PSRs B1641-45 and B0950+08 
\cite{paperIII}. 

    Wave growth in inhomogeneous media naturally results in bursty, variable waves. 
The field statistics are determined by the intrinsic statistics of radiation 
generated in an individual source region (predicted by theories for wave 
growth in inhomogeneous plasmas, which themselves depend on the emission 
mechanism and physics of the source plasma), effects of spatial variations 
due to possible superposition of emission from multiple (sub)sources, and 
scattering and other propagation effects between the observer and source, 
as reviewed in detail in Section 2 of Paper II. Only a very brief summary 
is given here, with most references provided in Paper II. First, the simplest assumption is 
made, that the observed field statistics are determined by the intrinsic statistics 
of the radiation process and propagation effects. Appeals to spatial variations 
are warranted only if the simplest assumptions fail to match data, which is not 
the case here. Second, scattering is expected to produce closely-Gaussian intensity 
statistics \cite{ratcliffe1956,rickett1977,paperII}, defined by equation (7) of Paper II 
and hereafter referred to as equation (II.7). Third, stochastic growth theory [SGT] 
\cite{rc2001}, corresponding to a linear instability operating near marginal stability, 
results in the lognormal statistics defined by (II.5). Fourth, self-organized 
criticality [SOC] \cite{baketal1987}, which corresponds to a 
strongly coupled system that is driven away from marginal stability but 
relaxes back via both local and system-wide events, has  power-law statistics defined 
by (II.6) with small indices. Fifth, for non-relativistic, weakly magnetized, 
electron-proton plasmas the nonlinear process of wave collapse \cite{r1997}, 
which involves the modulational (self-focusing) instability 
of a wavepacket, also results in power-law statistics but with higher indices 
that depend on wavepacket shape and effective dimensionality. Electromagnetic 
simulations for the relativistic, 
strongly magnetized electron-positron plasmas relevant to pulsars also show collapse 
occurring but the field statistics were not investigated (Weatherall 1997, 1998). Since 
collapse is qualitatively 
very similar in both sets of simulations, it is presumed below that collapse in 
pulsar environments also leads to power-law statistics with similar indices to those 
for electron-proton plasmas. Sixth, nonlinear 
decay processes, in which a primary wave decays into product waves, 
cutoff the field distribution with known functional form near and above the 
processes nonlinear threshold. Note that SGT can coexist at moderate fields with 
wave collapse or decay active at higher fields. 
%These predictions are 
%compared with the Vela pulsar's field statistics in this paper and in paper II. 

\begin{figure}
\psfig{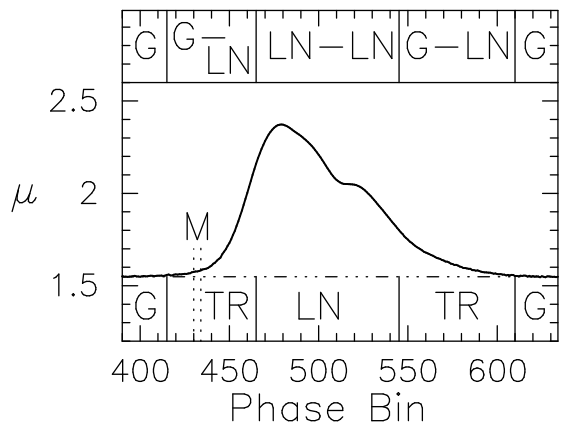}
\caption{Vela's average pulse profile as function of phase bin, where  
$\mu = \langle \log E \rangle$ and the horizontal dash-dot line is the off-pulse average. 
At bottom symbols and 
vertical lines identify the transition regions (TR) between regions where the 
field statistics are field statistics are approximately Gaussian (G) or 
lognormal [LN] (Papers I \& II). 
Vertical dotted lines and the symbol M show where giant micropulses occur  
(Johnston et al., 2001). At top symbols based on this work indicate where  
best-fits are Gaussian, 
Gaussian-lognormal (G-LN) or double lognormal (LN-LN).}
\label{profile}
\end{figure}

   Paper II demonstrates multiple results for the Vela pulsar. (i) 
The observed probability distributions $P(\log E)$ of the wave field $E$ are well-defined 
as functions of the 
pulse phase $\phi$. (ii) These distributions evolve with $\phi$, from Gaussian 
intensity statistics off-pulse, to approximately power-law statistics in a 
transition region, to lognormal statistics near the pulse center, and thence back 
through approximately power-law statistics to Gaussian intensity statistics off-pulse 
again. Figure \ref{profile} illustrates this behaviour. (iii) The off-pulse data 
are quantitatively consistent with Gaussian intensity statistics. 
(iv) The field statistics near the pulse center are 
consistent with the lognormal form (II.5) provided analysis is restricted to fields well 
above the noise background. (v) However, at low $E$ the observed $P(\log E)$ distributions  
lie consistently above the lognormal fit. This suggests that a second population 
of waves is superposed on the lognormal component. 

   Figure \ref{profile} also shows the restricted phase range ($\approx 430 - 434$) 
where Vela's giant micropulses are observed \cite{johnstonetal2001,krameretal2002}: 
defined by analogy with giant pulses \cite{lundgrenetal1995}, which have 
pulse-integrated fluxes exceeding $10$ times the average pulse-integrated flux,  
giant micropulses have fluxes greater than $10$ times the average flux at that 
phase. Giant micropulses and pulses are discussed further in Section 7. 

   Superposition of at least two wave populations is expected for the pulsar, since the 
Gaussian population observed off-pulse should extend on-pulse if it corresponds to 
measurement noise, sky background, or scattered radiation. Accordingly here we 
analyze the Vela data in detail in terms of 
two superposed wave populations, using a new theory for the statistics of two populations 
of transverse wave fields with known statistics that are vectorially combined together 
(Cairns, Robinson \& Das 2002c). This theory is summarized in Section 2. The dataset is described 
briefly in Section 3. In Section 4 we demonstrate 
in detail that the approximately power-law $P(\log E)$ distributions observed in the 
transition regions are consistent with vector convolution of a Gaussian intensity 
distribution with a lognormal distribution. These fits consider 
almost all available data and typically have very high statistical significance. 
Section 5 addresses the lognormal region, showing that the combination of two 
lognormals provides very good fits to the available data, being demonstrably 
superior to the single-lognormal fits in paper I and to Gaussian-lognormal combinations. 
Smooth evolution in the fit parameters across the source is demonstrated in Section 6, 
providing additional confidence in the fits and fitting procedure. Moreover, this 
smooth evolution also provides an additional argument that the approximately power-law 
statistics seen at phases in the transition regions are not intrinsic. It is argued in 
Section 7 that these analyses demonstrate that the Vela pulsar's field statistics are consistent 
with lognormal statistics, and so SGT, being 
relevant. The analyses are consistent 
with the pulsar's emission mechanism being purely linear (either direct or indirect) in 
these phase ranges, with no evidence for a nonlinear emission mechanism, 
consistent with our earlier analyses (Papers I \& II). The results are also 
set in a wider context, comparing them with those for the Vela pulsar's giant 
micropulses and with solar system radio and plasma wave emissions. The paper's 
conclusions are summarized in Section 7. 

\section{Statistics of Two Vectorially-Superposed Wave Populations}
 
    Suppose the observed field ${\bf E}$ is formed by vector addition of two fields 
${\bf E}_{1}$ and ${\bf E}_{2}$ that are transverse to 
the measurement plane, result from different wave populations or source regions, and have 
separate probability distributions 
$P_{i}(E_{i}^{2})$ for $i = 1,2$. The probability distribution $P(E^{2})$ is then defined 
in terms of the magnitudes $E_{1}$ and 
$E_{2}$ and the angle $\theta$ between ${\bf E}_{1}$ and ${\bf E}_{2}$ by \cite{cetal2002c}
\begin{eqnarray}
P(E^{2}) & = & 2\pi A \int dE_{1}^{2} \int dE_{2}^{2} \int d\theta\  P(E_{1}^{2})\ P(E_{2}^{2})\ 
\nonumber \\ 
 & \times & P_{\theta}(\theta)\ \delta( E^{2} - E_{1}^{2} - E_{2}^{2} - 2 E_{1} E_{2} \cos \theta)\ . 
\label{p_def}
\end{eqnarray}
Here $P_{\theta}(\theta)$ is the probability 
distribution for $\theta$, $A$ is a normalization constant, and all probability 
distributions are normalized by $\int dE_{i}^{2} P_{i}(E_{i}^{2}) = 1 = \int dE^{2} P(E^{2})$. 
The delta function enforces the vector addition ${\bf E} = {\bf E}_{1} + {\bf E}_{2}$. 

\begin{figure}
\psfig{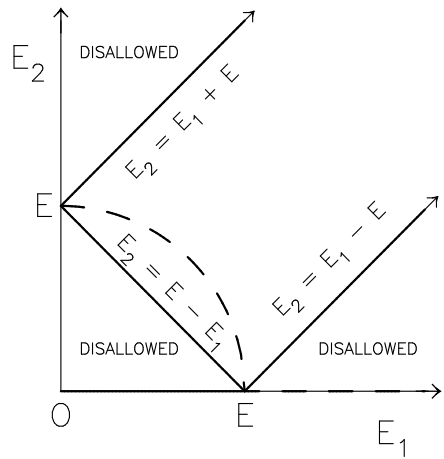}
\caption{Integration domain for (\ref{p_esq}), the lines of singularity associated 
with $\cos\theta = \pm 1$ (thick solid lines), and the circle corresponding to $\cos\theta = 0$ 
(dashed line). }
\label{p_esq_domain}
\end{figure}

   Assuming that $\theta$ is uniformly distributed, 
meaning that the two signals are produced independently in either the same source or different 
regions, $P(\theta) = (2\pi)^{-1}$). Then, integrating over the $\theta$ integral using the 
delta function leads to \cite{cetal2002c}
\begin{eqnarray}
P(E^{2}) & = & A \int dE_{1}^{2} \int dE_{2}^{2}\  
P_{1}(E_{1}^{2})\ P_{2}(E_{2}^{2})\   \nonumber \\
& \times & |\ [E^{2} - (E_{1} + E_{2})^{2}] [E^{2} - (E_{1} -
E_{2})^{2}]\ |^{-1/2}\ . 
\label{p_esq}
\end{eqnarray}
The square root in (\ref{p_esq}) contains integrable singularities corresponding to 
$\cos\theta = \pm 1$, where ${\bf E}_{1}$ and ${\bf E}_{2}$ are parallel or antiparallel. 
Figure \ref{p_esq_domain} shows the integration domain for (\ref{p_esq}), limited by the 
constraint 
$-1 \le \cos\theta \le 1$ and the physical portions of the singularities 
($E$, $E_{1}$, and $E_{2}$ 
must be positive-definite).  Two of the three singularity lines in Figure \ref{p_esq_domain} 
correspond 
to $\cos\theta = -1$ and the factor $E^{2} = (E_{1} - E_{2})^{2}$ in (\ref{p_esq}), implying that 
antiparallel fields ${\bf E}_{1}$ and ${\bf E}_{2}$ significantly affect the distribution 
$P(E^{2})$. The integral (\ref{p_esq}) is performed numerically. 

 The 
$P(E^{2})$ distribution is related to the corresponding intensity distribution 
$P(I)$ and the logarithmically-binned distribution $P(\log E)$ through their 
differentials and normalization 
conditions by 
\begin{equation}
P(E^{2}) = P(\log E) / (2 E^{2} \ln 10) = a P(I)\ . 
\label{relations}
\end{equation}
Explicit expressions for lognormal field distributions and for Gaussian 
intensity distributions are given in 
equations (II.5) and (II.7), respectively. 

\begin{figure}
\psfig{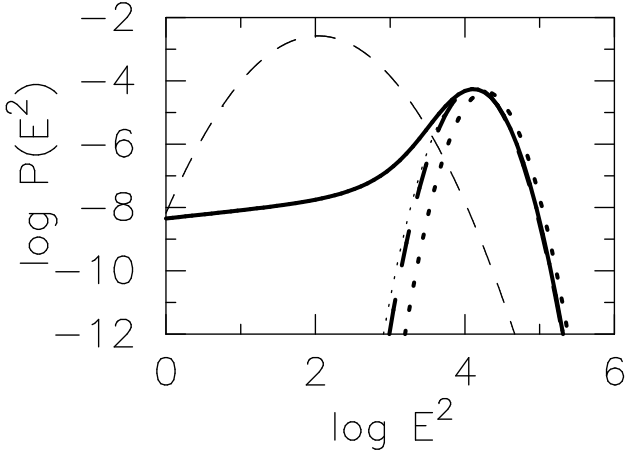}
\caption{Distribution $P(E^{2})$ resulting from vectorially combining two lognormal 
distributions: the thick solid line shows the combined distribution, while 
thick dashed and dotted lines show results for intensity and amplitude convolution, 
respectively, and thin 
dashed and dotted lines show the individual lognormal distributions. The 
thick dashed and thin dotted lines underlie the solid line at right.}
\label{p_tail}
\end{figure}

   Detailed analyses and explanations of the $P(E^{2})$ distributions that result from 
integrating (\ref{p_esq}) over the integration domain in Figure \ref{p_esq_domain} are 
described elsewhere \cite{cetal2002c}. Only four of their important qualitative results are 
summarized here 
and illustrated using the $P(E^{2})$ distribution predicted for two lognormal distributions 
in  Figure \ref{p_tail}. 

First, typically, as in Figure \ref{p_tail}, the combined 
$P(E^{2})$ distribution, at large $E$ above its peak, has the functional form of the 
individual distribution $P_{i}$ that extends to higher $E_{i}$. Expected intuitively when one 
distribution is much more intense then the other, this result explains why previous 
single-component analyses were usually viable. 

Second, a different functional form can 
be develop in the transition region between the two individual distributions. Specifically a 
nearly power-law form over $1-2$ decades can result from combining a Gaussian or 
lognormal distribution with a lognormal 
distribution that is centered at lower $E$ but extends to higher $E$ \cite{cetal2002c}. 

   Third, the prediction (\ref{p_esq}) for vectorial convolution of ${\bf E}_{1}$ and 
${\bf E}_{2}$ differs markedly from the corresponding predictions for convolution of the 
wave intensities or field amplitudes, as illustrated in Figure \ref{p_tail}. Accordingly, 
despite their familiarity and frequent use \cite{romanijohnston2001,johnstonromani2002}, 
intensity and amplitude convolution should not be used to interpret the detailed field 
statistics.  

Fourth, vectorial combination often produces a relatively flat 
tail at low $E$ below the peak of $P(E^{2})$. In Figure \ref{p_tail} this tail  
clearly does not come from the individual lognormal distributions, which show very rapid falloffs, 
but is instead due to overlap between the $P_{i}$ distributions for antiparallel vectors: since 
combining antiparallel vectors 
${\bf E}_{1}$ and ${\bf E}_{2}$ with similar magnitudes results in $E \approx 0$, the 
$P(E^{2})$ distribution will be significant at low $E$ if the 
$P_{1}$ and $P_{2}$  distributions overlap significantly. These 
features are shown below to be directly relevant to the Vela data. Specifically, the 
power-law distributions discussed in Section 3 are due to the second effect, the first result 
applies primarily in Section 4, and the tail at low $E$ is important observationally 
for all the Vela data. Note, via (3), the 
relatively flat tail for $P(E^{2})$ in Figure 
\ref{p_tail} becomes an $E^{2}$ power-law for the $P(\log E)$ distribution. 
Only $P(\log E)$ distributions are analyzed below.

\section{Data Set and Previous Fits}

   The Vela dataset analyzed is described in detail in 
Paper II and elsewhere (Johnston et al. 2001, Kramer et al. 2002), so only a minimal 
description is given here. The radiation fields on the Parkes antenna are 
detected as voltages, processed via a standard backend system, and then 
converted into calibrated fluxes $F$ averaged over the detector's 20 MHz bandwidth.
These flux samples are recast in terms of fields and intensities to permit 
direct comparisons between the data and theories for wave statistics. The 
variables $E'$ and $I'$ defined by 
\begin{eqnarray}
E' = & (F + I_{off}')^{1/2} & \propto E\ , \\
\label{E'defn}
I' = & F + I_{off}' & \propto I\ , 
\label{I'defn}
\end{eqnarray}
are related directly to the calibrated fluxes and are proxies for the 
incident field $E$ and intensity $I$ with units of 
(mJy)$^{1/2}$ and mJy, respectively. The telescope's backend system removes the 
time-steady levels of receiver noise, sky and supernova background, and pulsar emission. 
Accordingly, an offset $I_{off}' = 1250$ mJy, equal 
approximately to the rms noise level off-pulse (Johnston et al. 2001, Papers I \& II), 
is added to mitigate removal of the time-steady pulsar emission 
$\langle F_{psr} \rangle$, as justified 
in detail in paper II. Plausible causes for non-zero $\langle F_{psr} \rangle$  
are (i) coherently produced radiation that 
undergoes scattering and diffusion as it propagates from its source, perhaps 
changing the phase bin at which it is observed and producing a significant 
background, as well as (ii) synchrotron emission from the pulsar magnetosphere or 
jets. Detailed arguments against this procedure and the backend system 
significantly modifying the true 
statistics are presented in Sections 3 and 9 of paper II. Note that information 
on both microstructure and subpulses is retained.

Figure \ref{profile} above, Figure 1 of Paper I, 
and Figures 1, 3--5, and 12 of Paper II show the observational 
context in detail. As indicated in 
Figure \ref{profile}, and shown specifically in Figure 4 of 
paper II, the $P(\log E')$ distributions observed for phases $415 - 460$ 
and $545 - 610$ are approximately power-law at high $E'$. These distributions, in the 
so-called transition region, are 
fitted for the first time in section 4. 

   The $P(\log E')$ distributions observed in the approximate range $460 - 540$ 
appear strongly lognormal by eye and previous single-component lognormal fits are 
good and have reasonable statistical significance if 
restricted to bins with $\ge 10^{2}$ counts and fields 
$\gapprox 100$ (mJy)$^{1/2}$ (Papers I \& II). Typically these fits accurately 
model the observed distribution well at fields close to and above the 
peak in $P(\log E')$, but lie significantly below the observations  
at lower $E'$. This latter point argues for a second component contributing to the 
observed distribution \cite{paperII}, as shown in Section 4 for the power-law regions. 
Two-component double-lognormal fits for almost the entire range of $E'$ are presented 
for phases $460 - 540$ in Section 5.

\section{Two-Component Fits in the Transition Regions}

\begin{figure}
\psfig{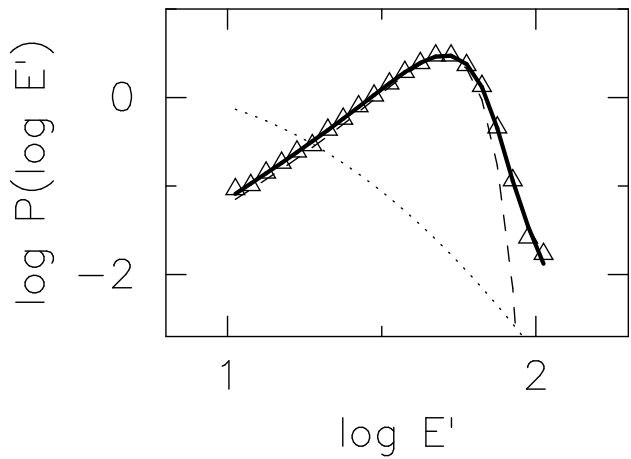}
\caption{Comparison of the $P(\log E')$ distribution observed 
at phase 440 (triangle symbols) with the best Gaussian-lognormal fit 
(thick solid line) to the prediction (\ref{p_esq}). The dashed line 
shows the Gaussian distribution that is vectorially convolved with the 
lognormal component (dotted line) to produce the best fit. }
\label{p_loge_440}
\end{figure}

   Figure \ref{p_loge_440} shows the $P(\log E')$ distribution observed at phase 440, 
the best fit to (\ref{p_esq}) for the combination of a Gaussian intensity 
distribution and a lognormal distribution, and the two individual distributions. The 
best fit was obtained by minimizing $\chi^{2}$ for all bins with $\ge 10$ counts using the 
amoeba algorithm \cite{pressetal1986}. In comparison, previous fits of similar Vela data 
(Paper I and II) considered only bins with $\ge 100$ counts and fields 
$E' \gapprox 100$ mJy$^{1/2}$ ($I' \gapprox 10^{4}$ mJy) and fitted only a single 
component. 

   Excellent quantitative agreement is evident in  
Figure \ref{p_loge_440} between the data and best-fit curve. The fit is very good statistically: 
Table \ref{table1} lists the fit parameters for the Gaussian component, $I_{0}'$ and 
$\sigma_{I'}$ in (II.7), and the lognormal component, $\mu$ and $\sigma$ in 
(II.5), as well as $\chi^{2}$, the number of degrees of freedom $N_{df}$, 
and the significance probability of the fit, $P(\chi^{2})$. Note that good fits 
have $\chi^{2} \approx N_{df}$ and $P(\chi^{2}) \gapprox 1\%$. 
Comparing the Gaussian fit parameters with the properties of the off-pulse (phases 
$391-400$) Gaussian intensity noise fitted in paper II, $I_{0}' = 1215$ 
mJy and $\sigma_{I'} = 1420$ mJy, it is clear that the Gaussian component found here 
corresponds closely to the off-pulse Gaussian background and to the offset $I_{off}'$ 
introduced in (\ref{E'defn}) -- (\ref{I'defn}). 

\begin{table}
\caption{Gaussian-lognormal fit parameters. $I_{0}$ and $\sigma_{I}$ are 
in units of mJy.}
\label{table1}
\begin{tabular}{ccccccccc}
\hline \\
Phase&$\mu$&$\sigma$&$I_{0}$&$\sigma_{I}$&$\chi^{2}$&$N_{df}$&$P(\chi^{2})$ \\
\hline \\
440&0.71&0.35&1804&1397&11.5&20&0.93 \\
450&1.13&0.40&2864&1285&33.9&26&0.14 \\
510&2.06&0.062&2633&2537&1551&15&0.00 \\
560&0.94&0.38&2474&1334&25.9&24&0.38 \\
590&0.68&0.40&1510&1284&18.1&21&0.65 \\
610&0.50&0.33&1248&1380&20.6&17&0.25 \\
\hline  \\
\end{tabular}
\end{table}

   The best-fit line in Figure \ref{p_loge_440} is very close to the underlying 
Gaussian distribution except for the $4$ data points with largest $E'$, where the data  
exceed the Gaussian predictions. This excess is due to the underlying lognormal 
distribution and corresponds to the highest $E$ signals from the lognormal adding to 
the Gaussian background. The simplest interpretation is that the lognormal distribution 
corresponds to the 
pulsar's intrinsic emission while the Gaussian component is a background signal. 
Evidence for evolution of the Gaussian background with phase is presented in Section 6 
below. Thus, for phase 440, the approximately power-law form below the peak of the 
observed $P(\log E')$ distribution corresponds to the Gaussian background while the 
upturn at high $E'$, which develops into a power-law at larger $\phi$, is due to the 
lognormal component.

\begin{figure}
\psfig{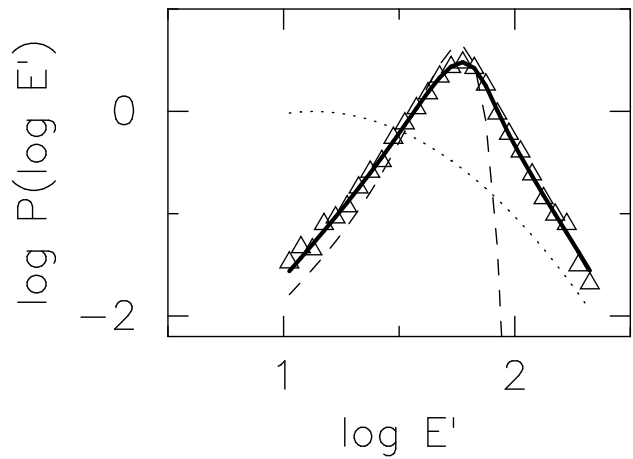}
\caption{Comparison of the $P(\log E')$ distribution observed 
at phase 450 (triangles) with 
the best Gaussian-lognormal fit (\ref{p_esq}), using the same format as Figure 
\ref{p_loge_440}.}
\label{p_loge_450}
\end{figure}

Figure \ref{p_loge_450} shows the observed $P(\log E')$ distribution and 
associated Gaussian-lognormal fits at phase 450, where the observed distribution 
has developed closely power-law forms at both low and high $E'$. For 
reference, $P(\log E') \propto E'\ ^{-4.1 \pm 0.5}$ at high $E'$ and 
$\propto E'\ ^{+3.1 \pm 0.3}$ at low $E'$ and at first sight the distribution appears 
rather similar to theory and observations for wave collapse discussed 
elsewhere \cite{rn1990,r1997}. Theoretical difficulties in interpreting these 
data in terms of collapse are described in Section 7. In contrast,  
the best-fit Gaussian-lognormal combination for (\ref{p_esq}) agrees  
very well with the observed distribution, with high statistical significance. 
Comparing Figures \ref{p_loge_440} -- \ref{p_loge_450} and the fit parameters in 
Table \ref{table1}, it is clear that evolution in the lognormal component explains 
simply the development of 
power-law character in the $P(\log E')$ distributions at high $E'$ for these phases. 
The ensuing examples show that this interpretation is also consistent for 
phase bins $< 440$ and in the other transition region (bins $545 - 610$).  
 
\begin{figure}
\psfig{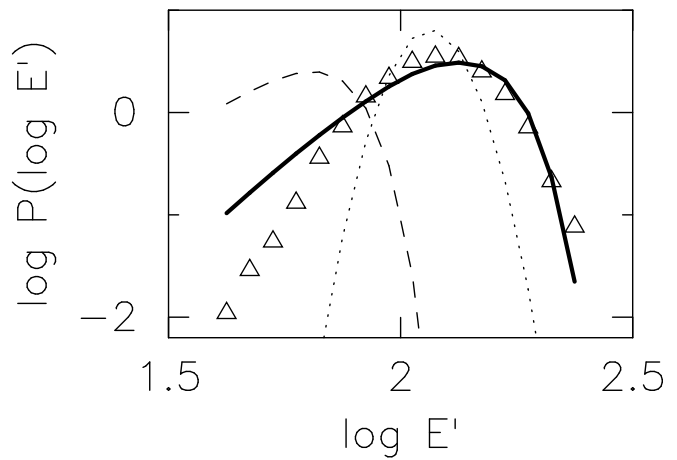}
\caption{Comparison of the $P(\log E')$ distribution observed 
at phase 510 (triangles) with 
the best Gaussian-lognormal fit, using the same format as Figure 
\ref{p_loge_440}.  }
\label{p_loge_gl_510}
\end{figure}

   Fitting the Gaussian-lognormal combination in the phase range $465 - 540$, where 
the $P(\log E')$ distributions look strongly lognormal as shown in Section 5 and in 
Papers I \& II, typically leads to worse fits and lower statistical significance 
than a double-lognormal combination. 
This is shown primarily in the next 2 sections, with Figure \ref{p_loge_gl_510} 
shown here mostly for consistency. While Figure \ref{p_loge_gl_510} shows reasonable 
agreement at high $E'$, the fit clearly fails for both the tail and 
the peak. 

\begin{figure}
\psfig{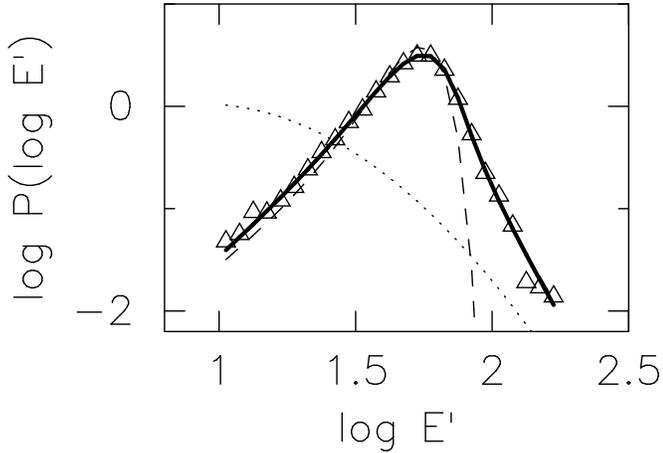}
\caption{Comparison of the $P(\log E')$ distribution observed 
at phase 560 with 
the best Gaussian-lognormal fit, using the same format as Figure 
\ref{p_loge_440}.  }
\label{p_loge_560}
\end{figure}

   Consider next the transition region in phase bins $545 - 610$ where the 
$P(\log E')$ distributions 
range from being initially power-law at high $E'$ towards being Gaussian in $I'$. 
Figure \ref{p_loge_560} shows the observed $P(\log E')$ distribution and associated 
Gaussian-lognormal fit at phase 560, where the distribution appears power-law both 
above and below its peak (indices $-6.8 \pm 0.9$ and $2.2 \pm 0.4$, respectively). 
Once again the agreement between the fit and data is very good, both quantitatively 
and in terms of statistical significance.   

\begin{figure}
\psfig{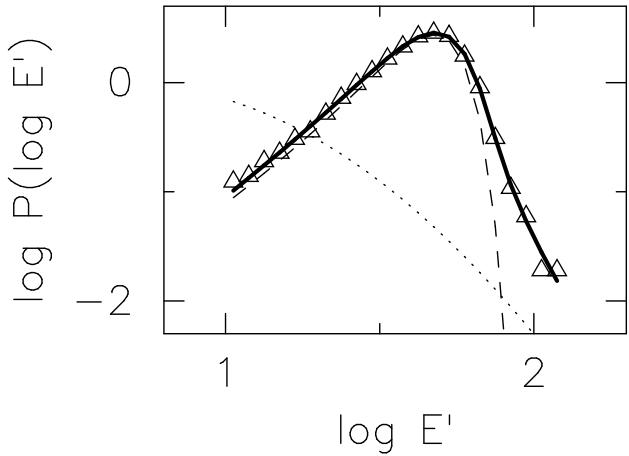}
\caption{Comparison of the $P(\log E')$ distribution observed 
at phase 590 with 
the best Gaussian-lognormal fit, using the same format as Figure 
\ref{p_loge_440}.  }
\label{p_loge_590}
\end{figure}

\begin{figure}
\psfig{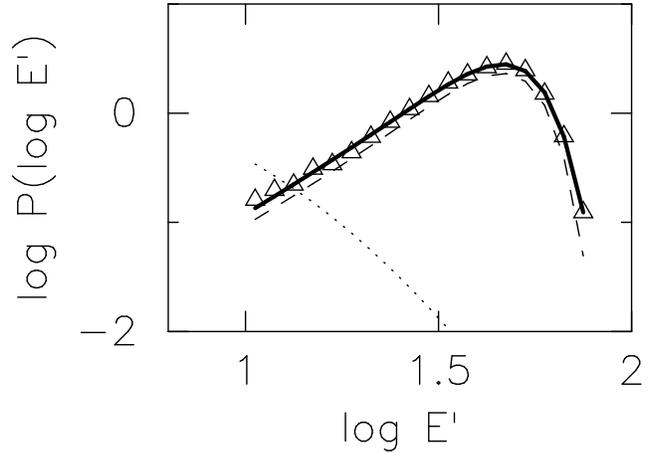}
\caption{Comparison of the $P(\log E')$ distribution observed 
at phase 610 (triangle symbols) with 
the best Gaussian-lognormal fit, using the same format as Figure 
\ref{p_loge_440}.  }
\label{p_loge_610}
\end{figure}

   As the pulsar's average intensity decreases for $\phi > 560$, and so 
the value $\mu$ of the inferred lognormal component decreases, 
the power-law feature at high $E'$ is expected to weaken and retreat into the 
background distribution by analogy with the results for phases 440 and 450. 
Figures \ref{p_loge_590} and \ref{p_loge_610} 
illustrate this trend for phase bins 590 and 610, respectively, and show that 
the observed $P(\log E')$ distributions
remain very well fitted by the Gaussian-lognormal combination. The lognormal 
component indeed has $\mu$ decreasing with increasing phase (Table \ref{table1}), 
while $\sigma$ remains 
approximately constant, showing that the lognormal moves towards and below the 
fields characteristic of the relatively constant Gaussian component. Similar 
results are obtained for phases 410 to 440, where the pulsar and lognormal 
component are moving above background.

%Figures \ref{p_loge_590} and 

   Phase 590 (Figure \ref{p_loge_590}) corresponds to the so-called 
``bump'' region \cite{johnstonetal2001} and shows that the bump region 
is not unusual, since the pulsar's emission continues to be well 
described as a lognormal component. Comparison with Figure 8a of Kramer 
et al. (2002)  emphasizes the importance of vectorial 
convolution of Gaussian and lognormal components via (\ref{p_esq}) 
for obtaining a good fit. 

In summary, the fits in Figures \ref{p_loge_440} to \ref{p_loge_610} and the 
associated variations in Table \ref{table1}'s fit parameters are strong observational 
evidence that the development of power-law-like features in the $P(\log E')$ distributions 
at high $E'$ in phase bins $415 - 460$ and $545 - 610$ correspond to combination 
of a Gaussian background component with an underlying lognormal component 
of the pulsar's emission. These power-law-like features are thus transitional stages 
between Gaussian intensity statistics off-pulse and strongly lognormal statistics 
near the pulse peak, justifying  the term ``transition regions'' 
adopted earlier \cite{paperII} for these phase domains. These data therefore 
provide no evidence of intrinsic power-law character for the pulsar's field 
statistics. In Section 7 below, it is shown that the power-law exponents for 
Figures \ref{p_loge_450} and \ref{p_loge_560} are inconsistent with collapse theory 
at low $E'$ and at best marginally consistent at high $E'$. These 
analyses are consistent with the dominant pulsar emission on-pulse being 
represented well by lognormal statistics. 

\section{Lognormal Region and Two-Component Fits}

   For most phase bins in the range $460 - 540$, fitting the 
observed $P(\log E')$ distributions with a Gaussian-lognormal 
combination leads to increased $\chi^{2}$ and worse fits, when the 
fitting algorithm converges at all, than for phases below 460 and 
above 540. Pursuing alternatives, we 
found that double-lognormal fits work very well for phases  
475 -- 510, with monotonically decreasing statistical 
significance from phases 475 to 450 and from 520 to 550. 
Adequate double-lognormal fits were not found at phases below $450$ or 
above $550$.  

\begin{figure}
\psfig{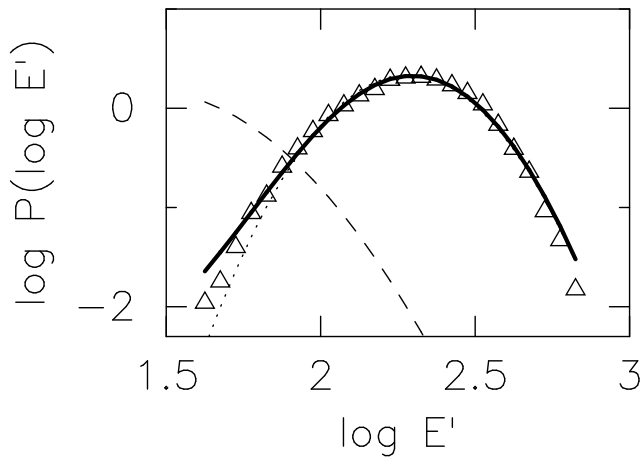}
\caption{Comparison of the $P(\log E')$ distribution observed 
at phase 470 (triangle symbols) with 
the best double-lognormal fit to (\ref{p_esq}),  using the same format as Figure 
\ref{p_loge_440} except that now both the dashed and dotted lines show 
lognormal distributions. }
\label{p_loge_470}
\end{figure}

Figure \ref{p_loge_470} shows that the  double-lognormal combination 
works well for phase $470$, which is very near the peak of the average 
profile (Figure \ref{profile}), fitting the observed distribution 
well except at very low and high $E'$. The fit parameters are listed in 
Table \ref{table2}. Note that the two individual lognormal distributions 
have very different parameters, with one centered at high $E'$ and 
the other at low $E'$ corresponding to the centroid $I_{0}' = E'^{2}$ in 
(II.7) of the off-pulse 
Gaussian component. This is why single-component fits restricted to 
high $E' \ge 10^{2}$ (mJy)$^{1/2}$ worked well in previous analyses 
(Papers I \& II). Similarly, at this phase fitting two components 
rather than one (ignoring only  
bins with $< 10$ counts) decreases $\chi^{2}$ from $188$ to $119$ but does 
not improve the fit's statistical significance qualitatively. 

\begin{table}
\caption{Double-lognormal fit parameters, with $\mu_{1}$ and $\sigma_{1}$ 
corresponding to the dominant component.}
\label{table2}
\begin{tabular}{cccccccc}
\hline \\
Phase&$\mu_{1}$&$\sigma_{1}$&$\mu_{2}$&$\sigma_{2}$&$\chi^{2}$&$N_{df}$&$P(\chi^{2})$ \\
\hline \\
470&2.28&0.18&1.49&0.26&119&24&$2\times 10^{-14}$ \\
480&2.37&0.095&1.62&0.22&5.4&15&0.99 \\
490&2.32&0.095&1.09&0.18&44.9&13&$2\times 10^{-5}$ \\ 
510&2.07&0.10&1.29&0.11&21.1&15&0.14 \\
540&1.86&0.043&1.38&0.14&373&18&$1\times 10^{-20}$ \\ 
\hline 
\end{tabular}
\end{table}

\begin{figure}
\psfig{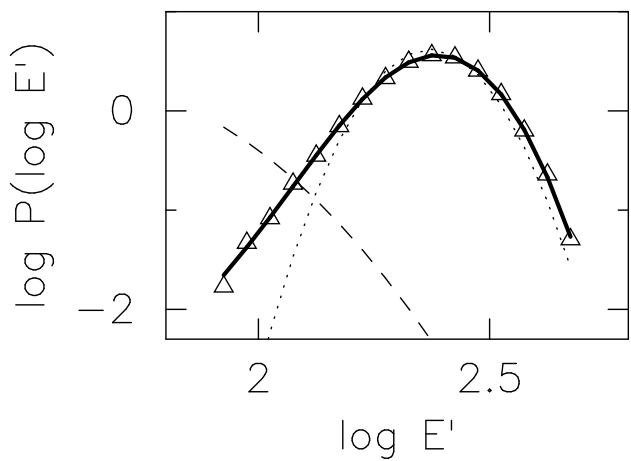}
\caption{Comparison of the $P(\log E')$ distribution observed 
at phase 480 with 
the best double-lognormal fit, in Figure \ref{p_loge_470}'s format. }
\label{p_loge_480}
\end{figure}

   The results for phase 480 are excellent (Figure \ref{p_loge_480}), with 
very high statistical significance for the fit and good agreement at all 
values of $\log E'$. Comparing Figures \ref{p_loge_470} and 
\ref{p_loge_480}, the range of the $P(\log E')$ distribution 
narrows significantly ($\sigma$ decreases) while its centroid ($\mu$) varies 
relatively little. This necessarily decreases the contribution of the 
off-pulse (Gaussian) component to the fit. Now $\chi^{2}$ decreases from 
$516$ to $5.4$ for the one- and two-component fits, respectively, corresponding 
to a qualitative change in statistical significance. A similar result is 
found at phase 475.

\begin{figure}
\psfig{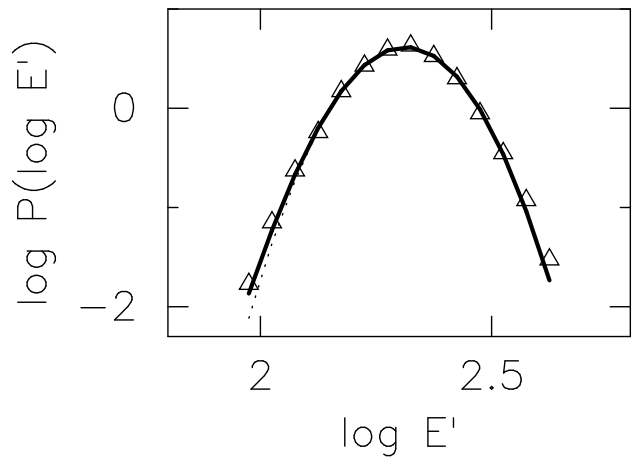}
\caption{Comparison of the $P(\log E')$ distribution observed 
at phase 490 with the best double-lognormal fit, in Figure 
\ref{p_loge_470}'s format. }
\label{p_loge_490}
\end{figure}

   Figure \ref{p_loge_490} presents the two-component fit for phase 490, 
showing it to agree very well with the data, both at large and small $E'$, 
with reasonable statistical significance. Comparing this with earlier 
single-component fits (Papers I \& II), the two-component fit agrees very 
well with the data over a much larger 
domain of $E'$.  

\begin{figure}
\psfig{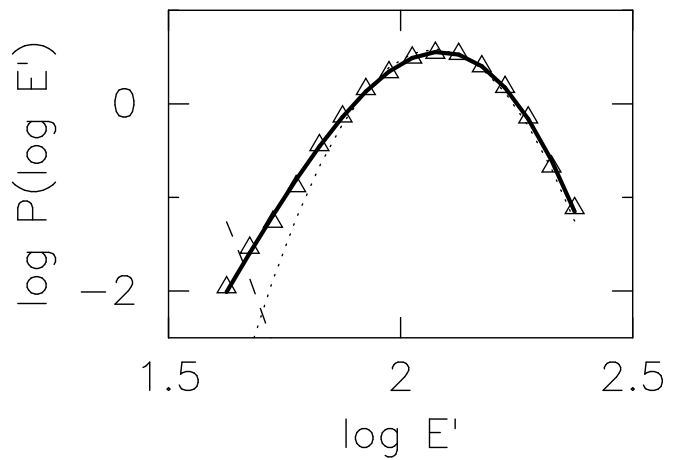}
\caption{Comparison of the $P(\log E')$ distribution observed 
at phase 510 with 
the best double-lognormal fit (\ref{p_esq}), using the same format as Figure 
\ref{p_loge_470}.  }
\label{p_loge_510}
\end{figure}

Figure \ref{p_loge_510} displays the double-lognormal fits for 
phase 510, which can be compared with the Gaussian-lognormal fit in 
Figure \ref{p_loge_gl_510}. The double-lognormal fits are clearly superior, 
correctly fitting the tail at low $E'$ and the peak of the distribution, which 
were missed by the Gaussian-lognormal 
fit. Tables \ref{table1} and \ref{table2} show the differing statistical 
significances of the fits. This case demonstrates that the nature of the individual 
distribution (e.g., Gaussian versus lognormal) is important and an optimum 
choice can be discerned from the fitting results, even when the difference 
is primarily important for only a few data points at low $E'$.  

%   A final example is for phase 540 (Figure \ref{p_loge_540}). 
%Compared with the corresponding single-component fit \cite{paperII}, the 
%double-lognormal fit agrees well at both high and low $E'$. The increasing 
%importance of the low-$E'$ component shows up in the extended tail at low 
%$E'$. At slightly greater phases Gaussian-lognormal fits become 
%better as Vela's average intensity and the fitted values of $\mu$ decrease, 
%and the $P(\log E)$ distributions develop power-law features at high $E'$ 
%as described in Section 4. 

%\begin{figure}
%%\psfig{file=p_loge_comb_540_ll.ps,angle=-90,height=6.0cm}
%\psfig{file=mnras_2_fig13_2.ps,angle=-90,height=6.0cm}
%\caption{Comparison of the $P(\log E')$ distribution observed 
%at phase 540 (triangle symbols) with 
%the best double-lognormal fit, using the same format as Figure 
%\ref{p_loge_470}.  }
%\label{p_loge_540}
%\end{figure}

\section{Evolution of the Fit Parameters With Phase}

    The Gaussian and lognormal fit parameters for the $P(\log E')$ distributions 
vary with the Vela pulsar's phase (Figure \ref{survey1}), as is evident also from comparing 
Figures 4 -- 14 and Tables 1--2. Values of $\chi^{2}$ less than or approximately equal to the 
number of degrees of freedom $N_{df}$ correspond to good fits from a statistical 
point of view. A number of important results are apparent: 

\begin{figure}
\psfig{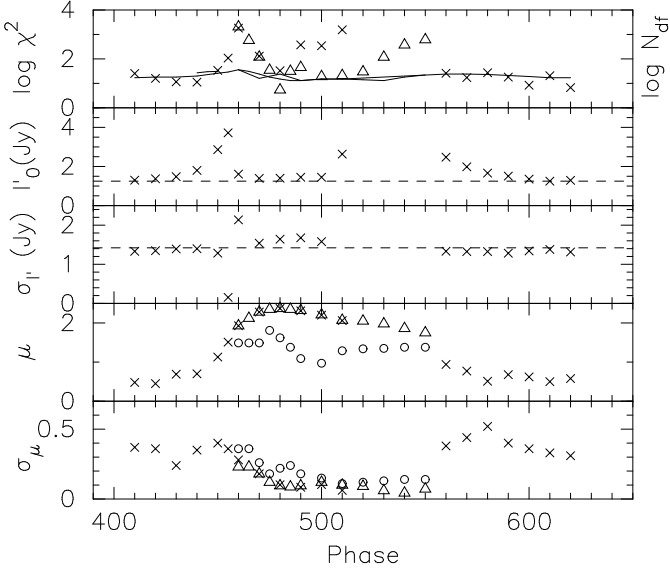}
\caption{Comparison of the fit parameters across the source. The top panel 
compares $\chi^{2}$  for the Gaussian-lognormal (crosses) and double-lognormal 
(triangles) fits with the number of degrees of freedom (solid lines). The 
second and third panels from the top show $I_{0}'$ and $\sigma_{I'}$ for 
the Gaussian-lognormal fits. The lowest two panels show the 
lognormal parameters for the Gaussian-lognormal fits (crosses) and 
the double lognormal fits (triangles and circles).}
\label{survey1}
\end{figure}

\begin{enumerate}
\item   Either the 
Gaussian-lognormal or double-lognormal combinations fit the observed $P(\log E')$ 
distributions well at essentially all phases. Phases where only one symbol 
appears indicates that either no fit was found for the other combination or 
else the other fit had $\chi^{2} \gapprox 10^{4}$. 

\item Double-lognormal fits are 
superior for $465 \lapprox \phi \lapprox 540$ where 
the pulsar's average profile peaks, while the Gaussian-lognormal combination is 
superior outside this domain. 

\item   The fit parameters vary 
smoothly with phase, except near phase bins $455$ and $555$ where the best-fit combinations 
are changing from Gaussian-lognormal to double-lognormal or vice-versa. Variations 
in the lognormal parameters are very large, consistent 
with the source plasma's parameters varying significantly with phase. 

\item The values of $\mu$ and $\sigma$ for the Gaussian-lognormal 
fits agree very well with those for the primary lognormal ($\mu_{1}$ and 
$\sigma_{1}$) from the double-lognormal 
fits. This shows that a high degree of confidence can be attached both to the fitting 
procedures and to interpretation of the pulsar's statistics and variability 
in terms of lognormal field distributions. 

\item Since the primary lognormal parameters vary smoothly where the 
superior fitting function switches between Gaussian-lognormal and double-lognormal, 
these transitions 
primarily correspond to the functional form of the low $E'$ wave distribution. 
This transition can be 
interpreted as evidence for evolution in the statistics of the less intense 
wave distribution. However, while these transitions are statistically significant, 
as implied by Figures \ref{p_loge_440} - \ref{survey1}, the interpretation is 
complicated by the fits yielding $I_{0}' 
\approx e^{2 \mu_{2}} \approx I_{off}' = 1250$ mJy (Figure \ref{survey1}). 
Here $\mu_{2}$ corresponds to the weaker lognormal for the double-lognormal fit. 
Any errors in 
$I_{off}'$ thus limit the quantitative (but not qualitative) significance of fit parameters 
$I_{0}'$ and $\mu_{2}$. 

\item The Gaussian component off-pulse and in the outer transition regions 
is best interpreted as background or 
measurement noise, which should be present at all phases. This is 
consistent with single-component Gaussian fits at off-pulse phases \cite{paperII} 
and the finding that $I_{0}' \approx I_{off}'$. There is weak evidence, however, 
that $I_{0}'$ increases from phase bins 430 to 455 and from bins 600 to 560, in the 
transition regions. This 
can be interpreted in terms of another Gaussian component dominating or adding to 
the off-pulse Gaussian 
component in these bins, perhaps being interpretable in terms of scattered radiation or 
another source of radiation with Gaussian statistics. Similar evidence for evolution of the 
Gaussian component across the source exists for other pulsars \cite{paperIII}. 

\item The transitions from Gaussian-lognormal to double-lognormal fits and vice versa 
can be interpreted as evidence for a third wave component 
appearing near the peak of Vela's 
average profile and dominating the off-pulse Gaussian component and the evolving Gaussian 
component possibly present in the transition regions. 

\item    Finally, the bottom panel of Figure \ref{survey1} shows that $\sigma$ 
is large near the 
pulse edges and small near the pulse center. This is consistent with earlier 
statements that pulsar modulation indices behave in the same way 
\cite{tayloretal1975,johnstonromani2002}; however, the new results place the 
earlier statements on a much stronger physical basis, since they 
demonstrate that the Vela pulsar's variability corresponds to lognormal 
statistics and that the lognormal parameter $\sigma$ varies in this way across 
the source, both significant steps towards developing a detailed physical model 
for the source plasma and emission physics. Such a model would link the wave 
parameters with instability physics and medium inhomogeneities, 
as done in several solar system contexts \cite{retal1993,cm2001}. 
\end{enumerate}
 
\section{Discussion}

  The existence of very good Gaussian-lognormal or double-lognormal fits to the 
observed $P(\log E')$ distributions at almost all on-pulse phases (excluding the 
giant micropulses discussed below) and the smooth evolution of the fit 
parameters with phase constitute very strong evidence that lognormal 
statistics, consistent with SGT, are relevant at all on-pulse phases for the 
Vela pulsar.  
Put another way, the pulsar's variability at all phases can be interpreted as the 
result of a pure SGT system coupled with a second group of weaker waves 
that have either Gaussian intensity statistics or else lognormal statistics. 
These results thus generalize and strengthen the conclusions of Papers I \& II. 

   As remarked in Section 4, some $P(\log E')$ distributions in the 
transition region (Figures \ref{p_loge_450} and 
\ref{p_loge_560}, but not Figures \ref{p_loge_440} and \ref{p_loge_610}), 
have approximately power-law form and appear strongly 
reminiscent of wave collapse at first sight. On detailed examination, as shown 
next, the distributions are 
inconsistent with existing collapse theory. First, existing theory predicts that 
below the peak in the $P(E')$ distribution the index $\beta$ (with 
$P(E') \propto E'^{\beta}$) should vary with the dimension $D$ as 
$\beta = 2D -1$ for isotropic, $2D - 3$ for prolate, and $1$ for oblate 
collapse \cite{rn1990,r1997}. Converting from the spectral index for 
$P(\log E')$ to that 
for $P(E')$ using $E'\ P(E') = P(\log E')$, the observed values
of $\beta$ are $2.1 \pm 0.3$ and $2.2 \pm 0.4$ for phase bins 450 and 560, 
respectively. Isotropic collapse theory predicts $\beta = 3$ and $5$, 
for $D =2$ and $3$, while prolate collapse predicts $1$ and 
$3$, respectively. The exponents observed at low $E'$ are thus inconsistent 
with collapse. Second, the indices predicted for high $E'$ above the peak in 
$P(E')$, with $P(E') \propto E'^{-\alpha}$, are $\alpha = D +2$, $D + 3$, and $2D + 1$ 
for isotropic, prolate, and oblate collapse. For $D = 2 - 3$, these 
predictions are the integers $(4, 5)$, $(5, 6$, and $(5, 7)$, respectively. 
The values observed for phase bins $450$ and $560$ are $5.1 \pm 0.5$ and 
$6.8 \pm 0.9$, respectively. While the predicted and observed ranges thus overlap 
for $D \ge 2$, simultaneous consistency requires oblate collapse 
with $D = 2$ for phase $450$ and $D = 3$ for phase $560$. One theoretical 
difficulty here is, intuitively, that strong magnetization of 
pulsar magnetospheres is expected to make $D$ less than $3$ (cf. 
Asseo et al. 1990, Asseo 1996, Weatherall 1998). Moreover, the field 
statistics for collapse in pulsar magnetospheres need to be 
investigated (cf. Weatherall1997, 1998). 

   The analyses in Section 4 and the smooth evolution in fit parameters in 
Figure \ref{survey1} thus show that the $P(\log E')$ distributions 
observed in the transition regions (phases 435 -- 460 and 545 -- 610) are 
best interpreted as the 
vectorial convolution of a lognormal component with a Gaussian intensity 
component and not as an intrinsic power-law component. SOC, 
modulational instability, wave collapse, and driven thermal waves are then 
inconsistent with the data. Moreover, these results show that the existence of a 
power-law trend in the 
high-$E'$ field statistics does not automatically 
provide evidence for SOC, modulational instability, etc. but may instead 
be evidence for an SGT wave population convolved vectorially with a 
second population distributed as a Gaussian in $I'$ or another 
lognormal (implying a second SGT wave population).  

   The results in Sections 4 -- 6 therefore provide no evidence for an 
intrinsic power-law tail or for a nonlinear cutoff in the observed $P(\log E')$ 
distributions at high $E'$, which might have corresponded to nonlinear self-focusing or 
decay processes, respectively. This confirms, strengthens, and generalizes to 
almost the entire on-pulse domain earlier results for Vela (Papers I \& II): the 
simplest interpretation is that 
the field statistics are consistent with the emission processes 
at these phases being purely linear (either direct or indirect) 
and inconsistent with nonlinear emission mechanisms like wave collapse 
\cite{asseoetal1990,asseo1996,weatherall1998}. More generally, since spatial and 
temporal variations across the source might be important, the observations 
constrain viable nonlinear mechanisms to yield lognormal statistics when averaged over 
the ensemble of emitting structures in the source. This constraint is strong. Wave collapse, for 
instance, yields power-law statistics when ensemble-averaged over a 
homogeneous source \cite{rn1990,r1997}, 
and so is implausible. 

   The origin of the second population contributing to the wave statistics 
on-pulse is of interest. Note that contributions from the background sky and 
supernova remnant are expected to be negligible due to removal of the DC 
offset. Where the second population has Gaussian intensity statistics, it is 
interpretable in terms of measurement noise (e.g., thermal receiver noise),  
scattering and diffusion of pulsar radiation, and (less probably) superposition 
of radiation from multiple subsources. The close similarity of the Gaussian's 
parameters in the outer transition regions with those off-pulse suggests that measurement 
noise is the most probable explanation there. Scattering is not ruled out, however, and 
appears to be an attractive way to interpret the increases in the 
Gaussian fit parameter $I_{0}'$ in the inner portions of the transition 
regions. 

    Where the second population is lognormal, it is interpretable
in terms of emission from a second linear SGT system, either via a second 
emission process if generated in the same source as the primary population or 
else emission via the same or a different process in a second source region. Since the 
second lognormal's fit parameters are so different from those for the primary 
lognormal component, being restricted to fields $E'$ close to the off-pulse 
background, these parameters may be contaminated by background effects. It might be 
tempting to suppose that only receiver noise need be considered, but the evidence 
for evolution in the Gaussian component for Vela (Section 6) and other pulsars 
\cite{paperIII} and the transitions from the best fitting function being 
Gaussian-lognormal to being double-lognormal 
(and vice versa) with phase argues against this. Finally, while the present 
analyses do not distinguish between subpulse and microstructure effects, analyzing 
both without artificial distinction, it could be that the primary lognormal 
component is primarily associated with microstructure physics while the second 
component is associated with subpulse effects, scattering, and 
measurement noise.

   The preceding analyses and comments specifically exclude Vela's phase bins 
430 -- 434, where Johnston et al. (2001) observed ``giant micropulses'' with 
fluxes greater than 10 times the average flux at those phases. These 
giant micropulses have a cumulative probability 
distribution of the flux $F$, $\int_{-\infty}^{F} dF P(F)$, that appears power-law in $F$ 
with index $-2.85$ with an associated uncertainty (by eye) $\approx \pm 0.3$ 
\cite{krameretal2002}. 
Converting into a $P(E')$ distribution using (\ref{E'defn}) yields 
$P(E') \propto E'^{-\alpha}$ 
with $\alpha = -6.7 \pm 0.6$ for Vela \cite{c2002a}. Similar results are found 
for other pulsars with known giant micropulses or giant pulses and, moreover, 
this index lies within the range $4 - 7$ predicted for wave collapse  but lies 
outside the range $0.5 - 3$ for known SOC phenomena \cite{c2002a}. Further work 
on collapse theory is thus recommended for these 
giant phenomena, along the lines mentioned above. However, SGT also
remains viable provided the power-law distributions result from vector 
convolution of multiple wave populations, as found in Section 4 for the Vela pulsar's 
normal emissions. 
%Accordingly, interpretations in terms of modulational 
%instability and wave collapse are attractive for giant micropulses and pulses
%\cite{c2002a}, changing the suggested detailed application (from normal pulses to 
%giant micropulses) but otherwise giving some encouragement for earlier 
%workers on these processes \cite{asseoetal1990,asseo1996,weatherall1998}. 

   The foregoing results point to a richness in possible emission mechanisms 
and source plasma characteristics for pulsars and, more generally, for solar 
system and other astrophysical sources. Specifically, SGT appears to 
apply widely (Robinson \& Cairns 2001, Papers I, II and references therein) 
to plasma waves and propagating radio emissions, including both Vela and other 
pulsars, but does not apply to all emissions.  
Examples include the giant micropulse and pulse phenomena discussed just above, 
Jovian ``S'' bursts \cite{queinneczarka2001}, which have power-law flux statistics 
with an index of $2.0 \pm 0.5$ that lies in the range expected for SOC, and 
solar decimetric spike bursts which likely have exponential statistics 
\cite{islikerbenz2001}. These results show that analyses of field 
statistics can strongly constrain the source physics and 
mechanisms of natural emissions. 

\section{Conclusions}

   The variability and emission processes of the Vela pulsar can be 
investigated and strongly constrained using the statistics of the observed 
time-varying flux. The strong evolution of the pulsar's 
field statistics with pulse phase \cite{paperII}, from Gaussian in the 
intensity off-pulse to approximately power-law in $E'$ in the transition 
regions, where the average pulse profile is increasing or 
decreasing, and then lognormal near the 
peak of the average profile, show that field statistics allow 
probing of the source characteristics. Moreover, together with 
single-component fits to the pulsar's field statistics (Papers I \& II), 
these variations are strong evidence that multiple wave populations are often  
superposed. 

   This paper presents detailed two-component fits 
and associated interpretations of the Vela pulsar's field statistics, analyzing microstructure 
and subpulse effects simultaneously, using a new prediction  
for the amplitude statistics of the vector sum of two transverse fields with 
known statistics \cite{cetal2002c}. Excluding Vela's giant micropulses, 
it is shown that the approximately power-law field distributions observed in 
the transition regions are very well fitted by vector convolution of a 
lognormal with a Gaussian distribution in the intensity, the Gaussian 
distribution has properties very similar to the Gaussian noise observed 
at off-pulse phases, and the evolution with phase of the high-$E'$ power-law form 
occurs as a result of the lognormal component moving relative to the 
background level. Accordingly, these power-laws are not intrinsic but are 
instead evidence for the primary pulsar emission having lognormal statistics. 
In the phase domain where the field statistics are strongly lognormal to the eye, 
double-lognormal fits are superior to Gaussian-lognormal fits, with significantly 
improved domains of applicability and statistical significance compared with a 
single-component lognormal fit. Moreover, the fit parameters for the 
lognormal and Gaussian distributions vary smoothly, but also significantly, with 
phase. Thus, at essentially all on-pulse phases the 
observed field statistics are very well fitted by the vector combination of 
a lognormal with either a Gaussian or a second lognormal. This second component 
may be interpreted as measurement noise, scattered radiation, or multiple 
superposed sources if it is Gaussian, or a second population of waves produced in a 
linear SGT system (either via a different mechanism in the same source or 
alternatively by either the same or a different mechanism in a different 
source region) if it is lognormal. Put another way, the Vela pulsar's 
emission above background is always well represented in terms of at least 
one lognormal. Accordingly, these results demonstrate in detail that the 
Vela pulsar's field statistics are 
consistent with SGT being relevant, and with the emission mechanisms being 
purely linear, whenever the pulsar is detectable above background. The pulsar's 
variability thus corresponds to well-defined field statistics that are 
consistent with emission via a linear plasma instability in an SGT state. 
Similar conclusions are reached elsewhere \cite{paperIII} for PSRs 
B1641-45 and B0950+08. 

   The finding that Vela's lognormal parameter $\sigma$ is large near the 
pulse edges and small near the center is consistent with earlier results for 
pulsar modulation indices \cite{tayloretal1975,johnstonromani2002}. These 
variations in $\sigma$ and also $\mu$ with phase will allowing future probing of 
the source plasma, particularly when a theory for the pulsar's SGT parameters $\mu$ and 
$\sigma$ is developed. 
Finally, the richness observed in the field statistics of pulsars and of 
solar system emissions, plus the important constraints placed here on pulsar physics,   
demonstrate the power and utility of analyzing field statistics of 
astrophysical sources. 

\section*{Acknowledgments}

The authors thank B.~J. Rickett for helpful conversations and 
the Australian Research Council for financial support.

\bibliographystyle{mn}

%\appendix

\label{lastpage}
\end{document}